\documentclass[
]{ceurart}

\usepackage{listings}
\begin{document}

\copyrightyear{2024}
\copyrightclause{Copyright for this paper by its authors.
  Use permitted under Creative Commons License Attribution 4.0
  International (CC BY 4.0).}

\conference{AI for Access to Justice Workshop – Jurix 2024, 
Masaryk University, Brno, Czechia }

\title{Spreading the Risk of Scalable Legal Services: The Role of Insurance in Expanding Access to Justice}

\author{Roee Amir}[%
email=ramir@alumni.harvard.edu,
]
\cormark[1]

\author{David Chriki}[%
email=david.chriki@alumni.stanford.edu,
]
\cormark[2]

\author {Harel Omer}[%
email=harelomer@nvidia.com,
]
\cormark[3]
\fnmark[1]
\cortext[1]{LL.M., Harvard Law School; M.D. Candidate, Hadassah-Hebrew University School of Medicine; former Research Fellow, Harvard Law School Program on Corporate Governance.}
\cortext[2]{JSM, Stanford Law School}
\cortext[3]{Senior Product Manager, Nvidia}
\fntext[1]{We are grateful to Nathan Hemmendinger, Yotam Berger, and the participants of the 2024 JURIX AI Conference at Stanford Law School for their insightful discussions, valuable suggestions, and thoughtful feedback on earlier versions of this paper.}

\begin{abstract}
  Liability insurance for AI-powered legal services offers a promising solution to two critical barriers in using AI to expand access to justice: mitigating catastrophic risk to individual users from inadequate advice and ensuring meaningful accountability when failures occur. Existing accountability mechanisms face significant challenges: tort liability frameworks encounter barriers including judgment-proof providers and costly information asymmetries, while current regulatory approaches revolve around human oversight requirements, creating cost and scalability barriers which limit access to justice.  This Article argues that an insurance-based framework offers a promising response to these challenges by distributing risks across users while establishing market-driven incentives for quality improvement through performance-based premiums. The Article proposes a comprehensive insurance model for AI legal services that establishes clear risk thresholds, streamlined compensation mechanisms, and continuous performance monitoring. Rather than attempting to eliminate all risks through restrictive ex-ante oversight requirements or relying on ineffective ex-post remedies, insurance enables efficient risk spreading while facilitating the scaling of automated legal services. This framework demonstrates how carefully structured insurance mechanisms can help realize AI's transformative potential to democratize legal assistance while maintaining robust user protections through sophisticated risk management rather than direct oversight. 
\end{abstract}


\maketitle
\section{Introduction }

Liability insurance for legal services powered by artificial intelligence (AI) offers a promising solution to two primary barriers in using AI to expand access to justice: mitigating the risk that individual users face catastrophic consequences from inadequate advice and ensuring accountability when such failures occur. While legal scholars, policymakers, bar associations, lawyers, and courts have emphasized responsible AI use — often advocating for limited scope or mandatory human oversight — these restrictions could severely constrain the technology's capacity to democratize legal assistance.\footnote{A number of jurisdictions ban unauthorized practice of law in a manner that could effectively limit the ability of companies to provide legal advice without oversight of licensed lawyers.\cite{bonardi2024certifying} \cite{avery2024chatgpt}. Even when the use of AI legal advisors is not strictly banned, courts have penalized unrepresented individuals for not correcting legal arguments in pleadings generated by generative AI, though the unrepresented individual does not possess the expertise to identify such errors\cite{kruse2024} } Insurance accomplishes these dual objectives by distributing potential losses across all users rather than concentrating risk on individuals, while simultaneously creating a reinforcing loop for accountability through performance-based premiums and deductibles that incentivize service providers to maintain high standards of accuracy. This approach preserves the transformative potential of AI legal services while protecting individual users and establishing clear financial consequences for inadequate performance.

The emergence of AI-powered legal services presents an unprecedented opportunity to expand access to legal services. Through sophisticated language processing and generation capabilities, these tools automate key legal processes that have traditionally required human lawyers, empowering individuals to articulate claims, engage in better-informed settlement negotiations, and navigate legal disputes. Such systems promise to bridge systemic gaps in legal service delivery by reducing costs, enabling scalable solutions, and addressing long-standing disparities in access to justice. 

However, the transformative potential of AI is shadowed by significant risks. AI-generated legal advice can be flawed or misleading, potentially exposing users to severe adverse consequences. This risk warrants particular attention for two fundamental reasons. First, the technical complexity and opacity of AI systems often leaves users ill-equipped to assess the reliability of the guidance they receive or make informed decisions about using these services despite the risks involved. Second, when these tools fail, the burden falls disproportionately on individual users, while the system's overall success in serving most users masks these failures. This statistical success may diminish providers' incentives to improve accuracy or fully disclose limitations, even as their errors impose catastrophic losses on individual users—creating a stark accountability deficit that threatens to erode trust and limit adoption of these promising services. 

Both traditional accountability mechanisms and current regulatory frameworks seem inadequate to address these challenges. Traditional tort liability faces several fundamental barriers: the technical complexity and opacity of AI systems make causation and fault exceptionally difficult to establish, rendering potential claims prohibitively expensive to litigate and leading to systematic underenforcement. Even if users could successfully establish liability, AI-powered legal services present an acute judgment-proof problem, where potential damages from systematic errors could far exceed providers' assets. Perhaps most critically, pursuing tort claims requires precisely the legal expertise and resources whose absence drives users to seek AI assistance in the first place. Meanwhile, current regulatory approaches focusing on mandatory human oversight of these technologies involve high costs and create scalability bottlenecks that undermine the technology's potential to democratize legal assistance.

This Article argues that liability insurance for AI-powered legal services offers a promising response to these challenges. Rather than attempting to eliminate all risks through constrictive regulatory measures or relying solely on traditional legal remedies poorly equipped to address AI systems' unique characteristics, insurance enables efficient risk spreading while facilitating the scaling of automated legal services. The proposed insurance framework would operate on multiple levels to address current shortcomings: First, by distributing potential losses across all users, it protects individuals from catastrophic consequences while providing streamlined paths to compensation through predetermined mechanisms rather than complex litigation. Second, the framework creates market-driven accountability through performance-based premiums and deductibles, creating direct financial incentives for accuracy and reliability. This accountability would be reinforced by insurance providers' own incentives to continuously monitor AI system performance against established risk thresholds, requiring services to maintain measurable standards of quality. This shift from direct oversight and case-by-case litigation to systematic risk management enables the development of scalable, accessible legal services without sacrificing user protection or stifling innovation.

The analysis proceeds as follows. Beyond this introduction as \textbf{Part 1}, \textbf{Part 2} examines the transformative potential of AI-powered legal services, highlighting their promise to expand access to justice alongside the risks of flawed advice and inequitable outcomes. \textbf{Part 3} addresses accountability deficiencies in AI systems, emphasizing challenges in fault attribution and the limitations of traditional tort liability and current regulatory frameworks focused on human oversight. \textbf{Part 4} proposes liability insurance as a more effective framework for managing AI risks, detailing its ability to redistribute liability, incentivize system improvements through dynamic premiums, and align innovation with user protection. It contemplates implementation pathways, from regulatory mandates to voluntary, market-driven adoption. \textbf{Part 5} concludes by briefly reflecting on the broader implications of insurance-based accountability for the future of AI legal services.

\section{The Promises and Perils of AI-Powered Legal Services }

Generative AI has the power to revolutionize access to justice by democratizing legal assistance in unprecedented ways. Through advanced language understanding and generation capabilities, AI can help bridge the persistent gap between legal needs and accessible services. Whether helping users understand their legal rights, prepare documents, or navigate procedures, AI technologies offer new pathways to make legal assistance more accessible and affordable. This transformative potential manifests in myriad ways. For instance, AI can enable people without legal backgrounds to effectively express their claims in legal terms and arguments.\cite{chien2024robots}\cite{chien2024generative} This transformation of lay descriptions into legal claims advances access to justice far beyond mere court access, as identifying and naming unperceived injustices is crucial to dispute transformation.\cite{felstiner1980emergence} When people of any background can simply describe their situation to an AI system and ask "do I have a legal claim," receiving clear guidance on potential legal claims helps identify and name injustices, accelerating dispute emergence and resolution that might otherwise remain buried. Second, AI can help level negotiation playing fields by addressing information asymmetries. Once an injurious experience is named and transformed into a grievance, parties would typically 'bargain in the shadow of law,'\cite{mnookin1979bargaining}, negotiating within their perceived legal framework and anticipated court outcomes. Information asymmetries regarding legal rights can significantly impact these negotiations, and AI-powered legal advisors may help balance them by providing accessible, immediate guidance to all parties, despite potential accuracy limitations. Finally, even without information asymmetries, disparities in litigation costs for initiating and maintaining formal legal claims may force parties with higher costs to accept reduced settlements — particularly when opposing parties have retained counsel.\cite{parchomovsky2012relational} AI-powered legal advisors could help mitigate such cost asymmetries, especially in early dispute stages, by providing affordable preliminary legal assessment. 

The market for AI-powered legal services is rapidly evolving in response to this transformative potential. The LegalTech sector is experiencing substantial growth, with projections indicating an expansion from \$29.60 billion in 2024 to \$68.04 billion by 2034.\cite{future_market_insights_legaltech_2024} Market actors across different sectors are recognizing and responding to this opportunity. For instance, major legal information providers are offering their generative AI tools to non-profit organizations to enhance their ability to serve clients in need,\cite{thomson_reuters_closing_2024} while specialized startups are developing targeted solutions for specific legal needs, from contract review to small claims court legal assistance.\cite{clover_contracts}\cite{peopleclerk} Beyond commercial developments, academic institutions and governments are actively engaging with AI's potential to expand access to justice, launching initiatives to study its implications and shape market incentives to promote innovation in service of this goal.\cite{stanford_ai_access_to_justice}\cite{texas_bar_meeting_minutes}\cite{bench_bar_of_minnesota}

While the potential benefits of AI-powered legal advisors are significant, courts, regulators, and scholars have focused considerable attention on the concern that they may provide flawed advice.\cite{bonardi2024certifying}\cite{jabotinsky2024ai}\cite{dahl2024large}\cite{statebar2023generative}\cite{kruse2024} However, the core policy challenges arising from the use of AI-powered legal services within the context of access to justice extend beyond the mere risk of errors. While the possibility of failures exists across all sectors and industries, two fundamental characteristics of this market make these risks particularly concerning and worthy of special attention.

First is the ability of users to make informed decisions about using these services despite the risks involved. While in theory, individuals facing legal challenges might rationally choose AI assistance over expensive counsel—or, when priced out of traditional legal services, over no assistance at all—weighing the potential risks against the cost savings. However, in the access to justice context, there's a substantial concern that this choice occurs under conditions of profound information asymmetry and bounded rationality that challenge standard assumptions of efficient risk allocation: AI-powered tools are primarily used by unrepresented individuals with limited legal knowledge, and the technical complexity and opacity of AI systems makes it extremely challenging for them to assess the quality of advice they receive, the probability of errors, and the potential damages they would incur if an error occurs. Even sophisticated users may struggle to properly evaluate AI-generated legal advice or accurately assess the magnitude of potential losses from reliance on incorrect guidance.\cite{harasta2024cannot} 

To illustrates this problem consider Joe, whose flight was canceled without justification, and who faces the airline's reluctance to reimburse the cost of the flight. Assume the flight tickets cost Joe \$50, the legal fees for standard legal advice would be \$500, and the airline retains a lawyer on a fixed cost, with an additional fee of \$10 per case. Under these circumstances, Joe would have little incentive to bring a claim against the airline, and would be inclined to settle for less than the expected value of the claim. If Joe can file a claim using an AI-based legal advisor for \$10, Joe will be more likely to pursue the claim and less likely to settle for any amount lower than the expected value of the claim. This remains true even if the AI advisor may provide bad advice, so long as the erroneous advice would not result in additional costs: as the chances of bad legal advice increase, Joe's chances to prevail in court decrease, and the expected value of his claim decreases accordingly, causing him to accept a smaller settlement. However, if courts might impose excessive costs on Joe for relying on AI-generated errors that include fictitious facts or case law, as in the case of \textit{Kruse v. Karlen},\cite{kruse2024} and Joe isn't aware of this risk, he might pursue a claim even when its expected value is negative—a decision he likely would not make if fully informed of the potential consequences.

Second, while AI-powered legal advisors may provide adequate advice in the majority of cases, when these tools fail, the burden falls disproportionately on individual users. This creates two distinct problems. First, this distribution of risk is fundamentally unjust: while both providers and the majority of users capture the economic benefits of these services' general reliability and scale, catastrophic losses fall entirely on individual users—often the most vulnerable individuals who lack resources to fully comprehend the potential risk and protect themselves from adverse outcomes. Second, the system's overall success in serving most users masks these individual failures, creating a classic market failure: although providers may generate reliable guidance in most instances, they have reduced incentives to improve accuracy or fully disclose limitations precisely because their statistical success obscures the devastating impact of individual failures. When users cannot effectively evaluate a platform's true error rates and limitations, providers lack sufficient market pressure to maximize reliability or transparency.\footnote{This market failure resembles Akerlof's 'Market for Lemons': just as used car sellers lack incentives to offer high-quality cars when buyers cannot verify quality before purchase, AI legal service providers lack incentives to maximize platform reliability when users cannot verify the true error rates and limitations of these systems \cite{akerlof1970market}.}

In summary, while AI-powered legal services show immense potential to reduce disparities in access to justice, significant structural challenges threaten to undermine this promise. The next section examines how current legal and regulatory frameworks attempt to address these challenges, revealing critical gaps that new approaches must bridge to realize AI's potential while protecting vulnerable users.

\section{The Accountability Challenge in AI-Driven Legal Services: Beyond Human Oversight}
The transformative potential of generative AI in legal services is undeniable. However, realizing this potential requires addressing a critical challenge: accountability. Unlike human actors in the legal system, AI systems cannot be held directly responsible for their actions. This lack of accountability raises fundamental questions about how to protect users from harm caused by flawed advice or system errors while preserving the innovation essential to improving and scaling these tools.

Accountability deficiencies in AI-powered legal services arise from the unique nature of these technologies. Traditional legal and regulatory frameworks are predicated on concepts of human agency, intent, and judgment to assign responsibility. These principles, however, do not readily apply to AI systems, which operate autonomously and often through opaque, "black box" processes. This disconnect creates a significant gap between the consequences of AI-generated legal advice and our ability to identify and assign fault. As a result, users are left vulnerable to harm, and trust in these tools is eroded, particularly among marginalized communities that stand to benefit most from their adoption. Courts, reflecting this uncertainty, have begun mandating disclosure of AI use and imposing penalties for errors—penalties that likely would not be sanctioned if such mistakes were made unintentionally by human lawyers.\cite{bonardi2024certifying}\cite{avery2024chatgpt}\cite{kruse2024} 

This chapter explores these accountability challenges in depth, focusing on the structural limitations of existing mechanisms, particularly tort liability and regulatory frameworks that require mandatory human oversight. It contends that while these frameworks may be well-suited for traditional legal services, they struggle to address the unique risks posed by AI-powered legal services. By examining these deficiencies, the chapter advocates for the development of alternative frameworks that balance innovation with user protection. Such rethinking is essential to ensuring that AI fulfills its promise of democratizing access to justice without sacrificing fundamental safeguards or compromising core values in legal service delivery.

\subsection{Limitations of Traditional Tort Regimes}
Tort liability has been widely advocated as a mechanism for directing provider incentives toward enhanced product safety and ensuring compensation for victims of defective products.\cite{PolinskyShavell2010} However, a closer examination reveals several fundamental challenges that undermine both deterrence and compensation goals while potentially stifling innovation in this emerging field, rendering traditional tort frameworks inadequate for governing AI-powered legal services.

At the threshold level, AI legal service providers frequently limit their potential liability through contractual provisions. Many providers incorporate broad liability waivers in their terms of service, framing their offerings as "legal information" or "self-help" tools rather than legal advice—even though their systems effectively perform advisory functions.\footnote{\textit{See, e.g.}, terms of service of CourtRoom5\cite{courtroom5_terms}, AskLegal Bot\cite{asklegal_bot_terms}, 1Law\cite{1law_terms}, DoNotPay\cite{donotpay_terms}, LawBotPro\cite{lawbotpro_terms}, and Legal Robot\cite{legal_robot_terms}.} The complexity of these terms of service creates additional barriers for users who lack the legal expertise to evaluate their rights and potential claims, effectively eliminating avenues for redress before substantive legal hurdles even arise.

Beyond these contractual limitations, the technical complexity and algorithmic opacity of AI systems create additional barriers to effective liability enforcement through traditional tort law.\cite{Wendehorst2020} The 'black box' nature of these systems, as well as the information asymmetries between injurers and victims vis à vis understanding system capabilities, performance, and limitations, means that proving causation and fault through conventional litigation, if possible at all, requires extensive technical expertise and costly discovery. The costs of establishing liability in individual cases often prove prohibitive, resulting in systematic underenforcement and failing to provide either adequate compensation to victims or sufficient deterrence of careless deployment.

Even if users overcome these barriers to establishing liability, a more fundamental problem emerges: AI legal services present a classic judgment-proof problem, arising when potential liability far exceeds providers' assets.\cite{shavell1986judgment} As Shavell explains, parties that cannot fully pay for damages will disregard any potential liability exceeding their assets, undermining both the deterrence and compensation goals of tort law.\cite{shavell2005minimum} The problem is particularly acute in the context of AI-driven legal services, which currently are often provided by early-stage companies with limited funds. Though these small, lean companies can deploy automated services to thousands of users with relatively low costs, their assets may not suffice to cover even a fraction of the damages that could be caused to their users if systematic errors occur. Knowing that their funds are limited, these companies likely do not fully internalize the potential damages they may cause to their users. 

Perhaps most fundamentally, pursuing tort claims requires precisely the legal expertise and resources whose absence drives users to seek AI assistance in the first place. This creates a paradox where those most likely to rely on AI legal services—individuals priced out of traditional legal representation—are least able to seek redress when these services fail. The technical complexity of proving causation and fault, combined with the resources required for litigation, means that users who turn to AI assistance due to cost constraints face nearly insurmountable barriers to compensation when these systems cause harm.\footnote{Further complicating potential claims by users, many AI-based legal service providers include liability-limiting provisions in their terms of service. These contractual provisions make it even more challenging for lay users to understand whether they might have a viable claim after suffering losses from system errors—such as when their case is dismissed or they incur additional costs due to procedurally deficient filings. The complexity of these terms of service provisions creates additional barriers for precisely those users who lack the legal expertise to evaluate their rights and potential claims. } 

In addition to enforcement challenges, tort liability poses a broader systemic concern: its potential to stifle innovation precisely where experimentation and development are most needed. Pure liability regimes risk fostering excessive caution, particularly where substantial uncertainty surrounds new technologies. Research indicates that traditional liability rules often lead to overdeterrence, discouraging potentially beneficial developments.\cite{GalassoLuo2022} This chilling effect is exacerbated by courts' reliance on industry customs as benchmarks for liability, creating what Parchomovsky and Stein term an "anti-innovation bias".\cite{ParchomovskyStein2008} This bias proves particularly problematic for AI legal services, where iterative development and controlled experimentation are essential to improving system accuracy and reliability. Providers may respond to this bias by avoiding potentially transformative solutions for expanding access to justice, fearing that departures from accepted practices will expose them to heightened liability.

\subsection{Limitations of Current Regulatory Regimes}
Like tort liability frameworks, existing regulatory approaches aimed at addressing the accountability challenges of AI-driven legal services, by offering ex ante safeguards and ensuring quality control, often fall short of achieving their intended goals.

Current regulatory approaches incorporate human oversight as a central safeguard. This requirement for human review appears across multiple touchpoints between AI legal services and their stakeholders. First, many legal service providers effectively mandate human oversight by requiring users to seek lawyer review and disclaiming liability for AI-generated advice. These providers often frame their services as mere "legal information" or "self-help" tools rather than legal advice, while stating that users are solely liable for any errors and must consult licensed attorneys for specific problems.\footnote{Examples include DoNotPay ("If you need advice for a specific problem, you should consult with a licensed attorney." (\href{https://donotpay.com/learn/terms-of-service/}{https://donotpay.com/learn/terms-of-service/})); JusticeAI ("while helpful, JusticeAI is not a substitute for professional legal advice. For in-depth assistance and complex matters, consult a trained lawyer" (\href{https://justiceai.io/}{https://justiceai.io/}); Aux.ai ("The Information (Legal resources, forms, and chat) … is intended for general informational purposes only and should be used only as a starting point for addressing your legal issues. ... It is not a substitute for an in-person or telephonic consultation with a lawyer licensed to practice in your jurisdiction about your specific legal issue, and you should not rely on such Legal Information." (\href{https://www.1law.com/terms-of-use/}{https://www.1law.com/terms-of-use/}).} Second, courts have begun imposing significant penalties for errors in AI-generated legal content, emphasizing the importance of human review.\cite{kruse2024}. Third, at the regulatory level, the provision of AI-based legal advice without licensed lawyer review may be considered unauthorized practice of law subject to sanction in certain jurisdictions.\footnote{\textit{See, e.g.}, \cite{bonardi2024certifying}\cite{jabotinsky2024ai}\cite{dahl2024large}\cite{statebar2023generative}. While Utah has been allowing for legal service providers to experiment as part of a regulatory sandbox, no "high innovation" service has been approved, and consequently, only a limited number of AI-powered services are participating in the program, typically with certain involvement of lawyers. \textit{See} \cite{utah_innovation_office}}  

However, this approach introduces significant problems. Most fundamentally, human review dramatically increases operational costs, undermining the cost-saving potential of AI technology—creating a paradox where the very mechanism intended to make AI safe renders it financially inaccessible to those who need it most. The requirement for human oversight also creates significant bottlenecks that prevent AI legal services from scaling effectively, a limitation particularly problematic given the vast unmet need for legal services among underserved populations. Moreover, human oversight may provide a false sense of security while introducing its own potential for error, as lawyers reviewing AI-generated content may become complacent or overwhelmed by volume, potentially missing critical errors. These regulatory approaches reflect a limited understanding of AI technology's potential and, while well-intentioned, risk codifying an inefficient and ultimately counterproductive approach to AI accountability. By mandating human review, regulators may inadvertently create barriers to innovation in automated legal services while increasing costs and reducing accessibility, ultimately perpetuating existing inequities in access to legal services.

Certain jurisdictions have begun experimenting with alternative approaches through "regulatory sandboxes"—controlled environments where AI-powered legal services can operate under modified oversight frameworks. These experimental spaces typically grant participating companies selective waivers from specific regulatory requirements to facilitate innovation while maintaining basic consumer protections. The European Union's AI Act explicitly endorses such regulatory sandboxes for AI development\cite{EU2024AIAct}, while several jurisdictions have implemented similar frameworks specifically aimed at expanding access to justice by relaxing unauthorized practice of law restrictions.\cite{bonardi2024certifying}. Utah's regulatory sandbox program represents a leading example in the United States, incorporating dozens of legal service providers,\cite{UtahStandingOrder15} though only a limited number currently leverage AI-powered solutions.\cite{utah_innovation_office} Such experimental frameworks demonstrate growing recognition that traditional regulatory approaches may impede the development of innovative legal service delivery models.

However, while regulatory sandboxes hold promise for accelerating innovation in AI-powered legal services by reducing regulatory hurdles, they present their own limitations and risks. These frameworks typically maintain civil liability exposure for service providers, and may even increase practical barriers for users seeking remedies, as regulatory waivers could complicate negligence per se claims that would otherwise arise from regulatory violations. Thus, sandboxes do not address—and may even exacerbate—the fundamental limitations of tort liability discussed earlier in this paper, risking leaving end-users with a diminished practical ability to seek compensation when these services cause harm.\footnote{This concern could be mitigated by imposing strict liability for damages caused by services provided by companies that participate in a regulatory sandbox. \textit{See, e.g.} \cite{truby2022sandbox}However, even this approach faces severe limitations, considering the limited ability of end-users of access-to justice services to bring claims against their providers, and the limited ability of many of the startups that provide these services  to compensate their users due to their limit resources, as discussed above.} To realize their full potential in expanding access to justice, regulatory sandboxes must be integrated with robust compensation mechanisms that fulfill both remedial and deterrent functions. These mechanisms would not only protect users from financial harm but also incentivize systematic improvements in service quality through continuous oversight and feedback. Such an integrated approach ensures that increased flexibility in service delivery does not come at the expense of user protection or broader system accountability.

To conclude, this chapter demonstrated that existing regulatory frameworks fall short in addressing the accountability challenges posed by AI-driven legal services. The next chapter examines how liability insurance, based on a risk redistribution approach, can offer a more effective solution by promoting adequate user protection and incentivization the careful and safe scaling up of these promising tools.

\section{Insurance as a Risk Management Framework for AI Legal Services}

Liability insurance tailored for AI-powered legal services offers a promising solution to the two main challenges identified in the previous chapters. First, that end-users, particularly those without legal training, have limited ability to critically evaluate AI-generated materials, assess the quality of advice, or understand the potential risks involved in relying on such advice; and second, that the inherent statistical nature of AI models means that while these systems may perform reliably overall, certain users will bear the full consequences of occasional erroneous advice—a result which is not only unjust, but also reduces providers' incentives to improve accuracy or fully disclose limitations.

As we explain in this chapter, insurance addresses these challenges through two complementary mechanisms. Insurance provides streamlined compensation paths for users harmed by system errors, redistributing the risks associated with AI-based legal services across a broad user base and shifting the burden from individual users to providers and their insurers. Moreover, through dynamic premiums and monitoring systems, insurance creates market-driven incentives for ongoing quality improvement, helping ensure that providers invest in enhancing their systems' reliability and transparency.

\subsection{Protecting Users Through Streamlined Compensation Mechanisms}
Liability insurance shifts and redistributes the risks associated with AI-powered legal services, providing a robust framework to protect users from potential harms arising from erroneous advice. By pooling risks across a broad user base, insurance mechanisms alleviate the burden on individual users, who typically lack the expertise to evaluate AI-generated guidance or assess the potential consequences of errors. Instead, service providers obtain liability coverage to protect against claims, while insurers compensate users for damages caused by system errors. This model not only reduces individual risk but also builds trust and accountability in AI legal services, enabling their broader adoption without compromising user protection.

Under this framework, insurance provides clear and accessible compensation mechanisms specifically designed for AI-driven legal services. When users receive incorrect advice leading to adverse outcomes, they can file claims directly with the insurer through standardized processes. The claims process is streamlined and user-friendly, avoiding the complexity that might deter users from seeking redress. Insurers evaluate whether errors fall within predetermined covered categories, assess the resulting harm, and provide compensation accordingly,\footnote{Such process will also include a dispute resolutions mechanism. A full discussion on this process is beyond the scope of this paper.} This straightforward approach ensures that individuals, particularly those from underserved populations, have viable paths to compensation without requiring extensive legal knowledge or resources.

This insurance-based approach resolves several key shortcomings of tort liability in addressing AI-generated harms. First, it eliminates the substantial procedural barriers associated with proving causation and fault in AI systems, which typically require costly technical expertise and complex discovery. Second, it addresses the judgment-proof problem by ensuring that compensation is not dependent on the financial resources of individual service providers—particularly crucial given that many AI legal services are provided by early-stage companies with limited assets. Finally, insurance simplifies redress by removing the need for litigation, offering users direct access to compensation without requiring the very legal resources whose absence drove them to seek AI assistance in the first place.

The shift to an insurance-based compensation framework represents a fundamental reconceptualization of risk management in AI legal services. By redistributing liability and creating accessible pathways for redress, it protects users from financial harm while fostering trust in these emerging tools. This approach ensures that accountability mechanisms are both effective and equitable, enabling AI-powered legal services to scale responsibly while maintaining robust user protections.\cite{baker2013regulation}

\subsection{Reinforcing Accountability Through Premium Adjustments and Deductibles}
We propose that liability insurance for AI-powered legal services be designed to not only protect users but also promote accountability and continuous improvement among service providers. This could be achieved through mechanisms such as dynamic premiums, deductibles, and performance monitoring, which would create financial incentives directly linking system reliability to provider costs.\cite{baker2013law}

Under this model, companies developing and deploying AI-powered legal services would obtain liability insurance coverage before offering their services to the public, creating financial incentives for providers to improve their accuracy. The insurance framework would establish clear, measurable risk thresholds for different categories of legal services.\cite{koessler2024risk} For example, insurers might set different acceptable error rates for document review versus legal research or drafting services, with insurance providers continuously monitoring AI systems' performance against these thresholds through automated auditing systems, random sampling, and analysis of user complaints and claims. These sophisticated monitoring systems would track AI performance in real-time, analyzing patterns in user feedback, monitoring task complexity and risk levels, and assessing the accuracy of AI-generated advice against established legal standards. When errors or potential risks are identified, insurers could require immediate corrective action from service providers or adjust coverage terms accordingly.\cite{lior2023innovating}
Insurance providers would cover direct financial losses from AI system errors, while service providers would bear responsibility through deductibles and potentially increased premiums. For instance, if an AI service's error results in a \$5,000 small claims case being dismissed, the insurance might cover the full amount for the user, with the service provider responsible for a \$1,000 deductible. The relationship between insurance premiums and system performance creates a powerful feedback loop: companies maintaining high accuracy rates and low claim frequencies would benefit from reduced insurance costs, while those with poor performance records would face higher operating costs.\footnote{Real-time monitoring to determine adequate premiums has been introduced recently into cyber security insurance policies.\cite{kauflin2024how} ("Old-line insurers seemed hopelessly out of touch, sending prospective customers forms asking such basic questions as whether they had antivirus software installed. The newcomers, by contrast, scanned potential customers' systems as a hacker might. Sometimes they required specific security upgrades before agreeing to insure them. ... Even after Coalition or At-Bay accept a customer, they keep scanning it and sending alerts to control both their own and clients' risk. ... If a customer insists on installing software that's notoriously breach-prone, ... At-Bay will threaten to double its insurance premiums."} This dynamic pricing model would function as a market-based regulatory mechanism, pushing companies to invest in quality improvements and risk mitigation strategies. Over time, this structure could help AI-based legal services gain user trust, allowing them to scale responsibly and expand access to justice without the need for restrictive oversight.

An insurance-based framework represents a more sustainable and scalable approach to accountability in AI-based legal services, one that better aligns with the technology's potential to expand access to justice while protecting user interests. As AI continues to evolve and integrate into legal practice, insurance mechanisms offer a path forward that balances innovation with accountability, ensuring that the benefits of AI-driven legal services can reach those who need them most without sacrificing essential protections for users. The framework's emphasis on continuous monitoring, dynamic risk assessment, and financial incentives creates a robust accountability system that can adapt and scale alongside technological advancement in the legal sector.


The proposed insurance-based approach to risk management in AI legal services represents a fundamental shift in how we conceptualize quality control for legal assistance. Rather than relying on individual attorney oversight—a model that inherently limits scalability—insurance mechanisms distribute risk across large user populations while maintaining robust protections for individuals. This approach particularly benefits those who currently lack access to legal services, as it enables the development of affordable, accessible legal assistance while ensuring that users have recourse when errors occur. The success of this framework depends on careful calibration of premiums, deductibles, and risk thresholds to create the right balance of incentives while maintaining accessibility—ultimately supporting the broader goal of expanding access to justice through AI-driven legal services. Through this insurance framework, AI-based legal services can achieve the scalability necessary to serve previously unrepresented populations while maintaining accountability for system performance. This approach recognizes that expanding access to justice requires not just making legal services available, but ensuring their reliability and establishing clear mechanisms for redress when errors occur. By integrating insurance mechanisms with careful risk threshold management, we can create a sustainable model for expanding access to justice through AI technology. 

This market-driven accountability framework represents a more sustainable approach than traditional oversight mechanisms. Rather than relying on costly human review that impedes scalability, insurance enables controlled experimentation while maintaining robust user protections. The interplay of premiums, deductibles, and monitoring creates a sophisticated system that aligns the interests of providers, insurers, and users—ultimately supporting the responsible scaling of AI-powered legal services while ensuring their reliability and effectiveness in expanding access to justice.

\subsection{Managing Uncertainty Through Graduated Implementation} 
The inherent unpredictability of AI systems presents a significant challenge for traditional insurance frameworks. Unlike conventional risks that can be assessed through historical data and actuarial analysis, AI-powered legal services operate in a rapidly evolving technological landscape where potential failure modes and error patterns may not be fully understood. However, this uncertainty need not preclude the development of effective insurance mechanisms. Rather, it calls for a graduated approach that begins with well-defined, bounded risks and evolves alongside our understanding of AI system behavior.

The key to managing this uncertainty lies in developing specific, measurable risk categories with clearly defined coverage limits. Instead of attempting to insure against all potential AI-related risks—an approach that would likely prove both impractical and prohibitively expensive—insurers can focus initially on concrete, quantifiable outcomes. For instance, coverage might begin with straightforward procedural errors like missed filing deadlines or jurisdictional mistakes, where both the occurrence of an error and its financial impact can be clearly established. As insurers gain experience with these basic categories, coverage could gradually expand to encompass more complex scenarios, such as strategic advice or legal analysis.

Premium structures can be designed to reflect this graduated approach, with rates adjusting based on empirical performance data rather than theoretical risk assessments. This dynamic pricing model would leverage emerging benchmarks for AI system performance, such as error rates in document review or accuracy in legal research tasks. These benchmarks, while still in early stages of development, provide concrete metrics against which to assess system reliability and adjust coverage terms accordingly. As the industry accumulates more data on actual failure rates and their associated costs, insurers can refine their pricing models and coverage limits to better reflect real-world risk profiles.

The implementation of such insurance frameworks should follow a similar pattern of gradual expansion. Initial deployments might focus on lower-risk applications with limited potential for harm, allowing insurers to build expertise in assessing and pricing AI-specific risks. This experiential learning would inform the development of more comprehensive coverage options, with insurers adapting their models based on observed patterns of system behavior and user outcomes. Through this iterative process, the insurance industry can develop increasingly sophisticated approaches to managing AI risk while maintaining sustainable coverage options for service providers.

\subsection{Pathways to Insurance Adoption}
The adoption of liability insurance in AI legal services could emerge through several pathways. First, regulatory frameworks might mandate insurance coverage as a prerequisite for providing AI-based legal services, similar to mandatory malpractice insurance requirements already in place for lawyers in multiple jurisdictions.\footnote{Jennifer Ip \& Nora Rock, Mandatory Professional Indemnity Insurance \& a Mandatory Insurer: A Global Perspective, LawPRO Magazine 10:2, at 10, 10-12 (2011), www.practicepro.ca/ wp-content/uploads/2017/06/2011-09-mandatory-insurance-global-perspective.pdf accessed 30 November 2022}  Second, adoption could be driven by providers' anticipation of potential liability, as courts might increasingly hold AI service providers accountable for system errors, making insurance a necessary risk management tool. Third, insurance could emerge as a market differentiator, with providers voluntarily adopting coverage to signal reliability and build user trust.\cite{stern2022ai} This market-driven pathway could be reinforced by regulations requiring providers to disclose whether they carry insurance, enabling users to make informed choices while avoiding the judgment-proof problem before engaging with a service.\footnote{This appears to be the prevalent policy choice within the US, with over half of the states requiring disclosure in some form. \textit{See} Leslie C. Levin, Lawyers Going Bare and Clients Going Blind, 68 Fla. L. Rev. 1281 (2016)} Under this scenario, insurance coverage could become a competitive advantage, as users would likely prefer providers offering clear protection mechanisms and compensation paths. Determining whether mandatory insurance requirements or disclosure obligations would be normatively desirable and which of these pathways—or combination thereof—will prevail is beyond the scope of this paper, but it is likely to influence the final structure, pricing, and impact of insurance on how AI legal services develop and contribute to expanding access to justice.

\section{Conclusion: Transforming Access to Justice Through Insured AI Legal Services}
The emergence of generative AI presents an unprecedented opportunity to address the persistent crisis in access to legal services. However, this technological breakthrough brings fundamental challenges of accountability and risk management that threaten to limit its transformative potential. Traditional approaches to ensuring accountability through mandatory human oversight, while well-intentioned, ultimately perpetuate existing barriers to access by maintaining high costs and creating scalability bottlenecks. The insurance framework proposed in this paper offers an alternative path that enables the responsible deployment of AI legal services while expanding access to justice.

Insurance mechanisms offer a sophisticated approach to managing the risks inherent in AI legal services without sacrificing their scalability advantages. Shifting from a prevention-focused model based on human oversight to a compensation-focused model based on insurance maintains robust user protections while enabling broader access to affordable legal services. This framework creates powerful market incentives for continuous improvement in AI system performance through dynamic premium structures and risk thresholds, while ensuring that individual users are protected from catastrophic losses through risk pooling.

The insurance-based approach addresses several critical challenges in the deployment of AI legal services. First, it resolves the fundamental accountability deficiency inherent in AI systems by creating clear mechanisms for compensation when errors occur. Rather than requiring each user to bear the full risk of potential AI errors, insurance distributes this risk across the entire user base while creating financial incentives for service providers to maintain high standards of accuracy. Second, by eliminating the requirement for routine human oversight, this framework dramatically reduces operational costs, making legal services accessible to populations currently priced out of the market. Third, the continuous monitoring and feedback mechanisms inherent in insurance systems create a data-driven approach to quality improvement that can identify and address systematic issues before they result in widespread harm. The insurance framework enables the scaling of AI legal services while maintaining appropriate safeguards for different types of legal work. Through carefully calibrated risk thresholds and premium structures, insurers can create appropriate incentives for different categories of legal services, from routine document preparation to more complex advisory functions. This nuanced approach allows for innovation in serving traditionally underrepresented populations while maintaining rigorous standards for system performance.

The implementation of these insurance mechanisms requires careful attention to establishing appropriate risk thresholds and monitoring systems. However, the potential benefits are substantial: a new generation of AI-driven legal services that can provide reliable, accountable assistance at scale, without requiring constant human oversight. This transformation could dramatically expand access to justice, enabling millions of currently undeserved individuals to effectively assert their legal rights and navigate the legal system. The development and refinement of these insurance mechanisms represents a crucial step toward creating a legal system that is truly accessible to all. The insurance framework offers a practical pathway to harness the power of AI while ensuring appropriate protections for users. By enabling the responsible automation of legal services through insurance rather than oversight, this approach begins to address the persistent crisis in access to justice that has long plagued the legal system.

The future of legal services lies not in choosing between human oversight and automation, but in developing sophisticated mechanisms to manage risk while maximizing access. Insurance provides such a mechanism, offering a blueprint for the responsible deployment of AI legal services at scale. The insurance-based framework presented here demonstrates how the transformative potential of AI in law can benefit society as a whole through expanded access to justice, while maintaining robust protections for individual users through sophisticated risk management mechanisms.

\bibliography{bibfile}

\end{document}